\documentclass[aps,prl,twocolumn,superscriptaddress]{revtex4-1}

\usepackage{graphicx}
\usepackage{amsmath,amsfonts,amssymb}
\usepackage[colorlinks=true,linkcolor=blue,citecolor=blue]{hyperref}

\usepackage[separate-uncertainty = true,multi-part-units=single]{siunitx}

\usepackage{booktabs}
\usepackage{tabularx}

\begin{document}

\title{Role of hexagonal boron nitride in protecting ferromagnetic nanostructures from oxidation}


\author{Simon Zihlmann}
\affiliation{Department of Physics, University of Basel, Klingelbergstrasse 82, CH-4056 Basel, Switzerland}

\author{P\'eter Makk}
\email{peter.makk@unibas.ch}
\affiliation{Department of Physics, University of Basel, Klingelbergstrasse 82, CH-4056 Basel, Switzerland}

\author{C. A. F. Vaz}
\affiliation{Swiss Light Source, Paul Scherrer Institut, CH-5232 Villigen PSI, Switzerland}

\author{Christian Sch\"onenberger}
\affiliation{Department of Physics, University of Basel, Klingelbergstrasse 82, CH-4056 Basel, Switzerland}

\date{\today}
\begin{abstract}

Ferromagnetic contacts are widely used to inject spin polarized currents into non-magnetic materials such as semiconductors or 2-dimensional materials like graphene. In these systems, oxidation of the ferromagnetic materials poses an intrinsic limitation on device performance. Here we investigate the role of  ex-situ transferred chemical vapour deposited hexagonal boron nitride (hBN) as an oxidation barrier for nanostructured cobalt and permalloy electrodes. The chemical state of the ferromagnets was investigated using X-ray photoemission electron microscopy owing to its high sensitivity and lateral resolution. We have compared the oxide thickness formed on ferromagnetic nanostructures covered by hBN to uncovered reference structures. Our results show that hBN reduces the oxidation rate of ferromagnetic nanostructures suggesting that it could be used as an ultra-thin protection layer in future spintronic devices.

\end{abstract}

\maketitle

\section{Introduction}
\label{sec:Introduction}
Spintronics is a growing field where the injection, detection, active control and manipulation of spins give the basis of solid-state electronic circuits \cite{2004_Zutic}. Therefore, the creation of spin polarized currents in a non-magnetic material, which is most commonly achieved by connecting ferromagnetic contacts to it, is the cornerstone for any study on the spin properties of the host material. It has been recently found that 2D-materials offer a new platform for spintronics devices, owing to their wealth of unusual physical phenomena and great diversity \cite{2014_Han}. Graphene is a promising candidate for a spin channel material since it has a low spin orbit interaction and nearly no nuclear spins, resulting in long spin relaxation times \cite{2014_Han, 2012_Seneor}. Other 2D materials, with higher spin-orbit coupling, or their combination with graphene could be used for spin manipulations.

Oxide layers are commonly used in modern spintronics devices as tunnel barriers.  In magnetic tunnel junctions, which probably are the most widely used spintronics devices, the oxide layer is the key ingredient for achieving large signals \cite{1995_Miyazaki, 1995_Moodera}, whereas in spin-valves they are used for circumventing the conductivity mismatch in spin injection from metallic structure into semiconductors or graphene \cite{2000_Schmidt}. However, high quality oxide tunnel barriers are hard to grow on 2D materials (e.g. graphene) \cite{2008_Wang, 2008_Wang_a}. It has only been recently that hexagonal boron nitride (hBN) with a band gap of \SI{6}{eV} \cite{2004_Watanabe} has been used as a crystalline pin-hole free tunnel barrier \cite{2011_Lee, 2012_Britnell, 2012_Amet, 2015_Chandni}. Spin injection into graphene with exfoliated hBN \cite{2013_Yamaguchi} as well as with chemical vapour deposited (CVD) hBN \cite{2014_Fu, 2014_Kamalakar} tunnel barriers was also established. Furthermore, ferromagnetic / hBN heterojunctions were predicted to exhibit \cite{2009_Yazyev} and have shown \cite{2015_Dankert} large magnetoresistance.

Oxidation of ferromagnetic material is a challenging problem in spintronics. In  commercial devices, the ferromagnetic layers are always protected from oxidation by a layer of a noble metal (e.g. Ru). It has been shown that even a single layer of graphene is enough to protect a Ni electrode from oxidation \cite{2015_Martin}. However, an insulating coating would have the advantage that it could act as a tunnel barrier for spin injection and as an oxidation barrier at the same time. Recent experiments indicate that hBN could serve as an atomically thin oxidation barrier for nanostructured metallic contacts since it was already successfully shown for larger areas (macroscopic) and multiple hBN layers \cite{2013_Liu, 2014_Li}. Combined with advantages for spin injection, as mentioned above, hBN seems to be the perfect candidate to fulfill this dual role.

In this study we investigate the ability of hBN to act as an oxidation barrier for ferromagnetic nanostructures that could be used for electrical spin injection. Our measurements show that hBN can be used as a protective coating for metallic ferromagnetic materials. As a characterisation tool, X-ray photoemission electron microscopy (XPEEM) was chosen. XPEEM allows to investigate the chemical state (especially the oxidation) as well as the magnetic properties of the ferromagnetic material. In contrast to spatially integrated X-ray photoelectron spectroscopy or X-ray absorption spectra (XAS) methods, XPEEM offers the advantage of high lateral resolution of about \SIrange{50}{70}{nm} \cite{2012_Cheng}. Previously, XPEEM was succesfully used to determine the oxide thickness of ferromagnetic materials \cite{2001_Regan, 2014_Vaz}.

\section{Methods}
\label{sec:Methods}
\begin{figure}[htbp]
	\includegraphics[scale=1]{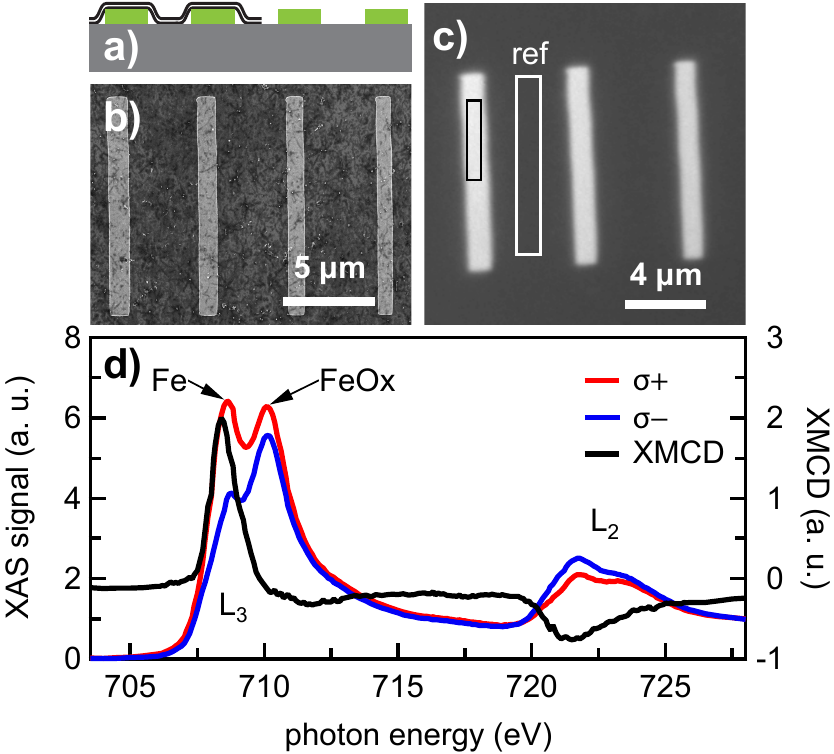}
	\caption{\label{fig:device}\textbf{Sample design and XMCD spectra}. \textbf{a)} shows a schematic cross section of the samples with the silicon substrate in gray, the ferromagnetic structures in green and the hBN as black solid lines. In \textbf{b)} a scanning electron micrograph of Py strips covered with hBN is shown. \textbf{c)} PEEM image of Py strips without hBN at the L$_3$ resonance of Fe. For further data analysis, spectra (black rectangle) and background spectra (white rectangle) were extracted. In \textbf{d)}, we show an XAS measured with circular polarized light (blue and red curve, left axis) and the corresponding XMCD signal (black, right axis). The L$_3$ edge shows a double peak with a magnetic contrast only on the left peak which is due to the metallic iron. The right peak at the L$_3$ edge shows only weak magnetic contrast since it is mostly due an antiferromagnetic iron oxide. As expected, the magnetic contrast is inverted at the L$_2$ edge compared to the L$_3$.}
\end{figure}

A schematic sketch of the samples investigated in this study is shown in Fig.~\ref{fig:device}~\textbf{a)}. Nanostructured ferromagnetic strips were fabricated by standard e-beam lithography on a Si substrate. After metallisation, half of the strips were covered with a bilayer (BL) of hBN, which was obtained from Graphene Supermarket. In \textbf{b)} we show a scanning electron micrograph (SEM) of permalloy (Py = Ni$_{80}$Fe$_{20}$) strips covered with hBN. Having both hBN covered and uncovered regions of ferromagnetic strips on the same sample allowed for direct comparison. Further details on the fabrication procedure can be found in the supporting information.

XAS were recorded at the SIM beamline at the Swiss Light Source. Linearly polarized photons with polarization axis perpendicular to the strip axis in grazing incidence were used for symmetric XAS in order to exclude any magnetic contrast. Circularly polarized photons were used for magnetic contrast imaging, probing the magnetization along the strip axis (easy axis of the nanomagnets) by taking advantage of the x-ray magnetic circular dichroism (XMCD) effect. The spectra were recorded by measuring the local intensity of photoemitted secondary electrons using XPEEM.

PEEM images were recorded as a function of photon energy. XAS can then be extracted at any point in these images. An example of a PEEM image is shown in Fig.~\ref{fig:device}~\textbf{c)}, where the Fe containing Py strips appear much brighter than the surrounding background. This image was recorded at the L$_3$ edge of iron \mbox{(E $\approx$ \SI{709}{eV})}. Here, the XAS of the ferromagnetic strips were extracted by averaging over a region on the strips (see black rectangle in Fig.~\ref{fig:device}~\textbf{c)}). The spectrum is then normalised by dividing the signal from the ferromagnet by the background signal (white rectangle in Fig.~\ref{fig:device}~\textbf{c)}) for every energy in order to compare the XAS from different samples. Furthermore, the spectra is then rescaled such that it is zero at the pre-edge ($\sim$~\SI{705}{eV}, no absorption) and one at the post-edge ($\sim$~\SI{727}{eV}, finite non resonant absorption), see also Fig.~\ref{fig:device}~\textbf{d)}. This normalization procedure makes the XAS from different samples directly comparable.

For each sample two independent regions were investigated (several hundred \si{\micro\meter} apart from each other). Four different regions on three different strips were used to extract the XAS signal (as in Fig.~\ref{fig:device}~c)) and the average of these four spectra was then used to extract the oxide thickness.

In Fig.~\ref{fig:device}~\textbf{d)} XAS at the Fe edge are shown for $\sigma+$ and $\sigma-$ polarized light in red and blue respectively. There are two main peaks corresponding to the spin-orbit splitting of the 2p core level, the L$_3$ edge (\mbox{E $\sim$ \SI{709}{eV}}) and L$_2$ edge (\mbox{E $\sim$ \SI{722}{eV}}). The two circularly polarized photons probe different transition probabilities into the spin-split 3d band and give rise to a magnetic contrast, which is given by the difference of the two XAS and is called XMCD: $(\sigma^+ - \sigma^-)$. The XMCD signal (black line) is positive at the L$_3$ edge and negative at the L$_2$ edge as expected since the average spin of the probing electron is inverted at the L$_2$ edge compared to the L$_3$ edge \cite{1995_Stoehr}. 
Two sub peaks at the L$_3$ edge are observed in the $\sigma^+$ as well as in the $\sigma^-$ XAS signal. A large magnetic contrast is observed for the left peak, whereas the right peak only shows very weak magnetic contrast. Metallic iron gives rise to a strong magnetic contrast due to its ferromagnetic nature and therefore we ascribe the left peak to the Fe peak (E~$\sim$~\SI{709}{eV}). Iron can form many different oxides, FeO, Fe$_3$O$_4$, $\alpha$-Fe$_2$O$_3$ and $\gamma$-Fe$_2$O$_3$ for example. FeO and $\alpha$-Fe$_2$O$_3$ are antiferromagnetic and will not show any magnetic contrast. Fe$_3$O$_4$ as well as \mbox{$\gamma$-Fe$_2$O$_3$} are ferrimagnetic and can contribute to a magnetic contrast, depending on the coupling to the ferromagnet below. Since $\gamma$-Fe$_2$O$_3$ has a spectral signature similar to Fe$_3$O$_4$ \cite{2011_Jimenez-Villacorta} it is difficult to distinguish from the latter. Therefore, we do not discriminate between the two here. Furthermore, the XMCD signal in Fig. \ref{fig:device}~d) shows a different line shape than expected for Fe$_3$O$_4$ and $\gamma$-Fe$_2$O$_3$ \cite{2004_Huang, 2006_Kim}. We conclude that the left peak can be ascribed to metallic Fe and the right peak (E~$\sim$~\SI{711}{eV}) can be ascribed to the iron oxides and therefore we will call the right peak the iron oxide peak.

\section{Results}
\label{sec:Results}
In order to investigate hBN as an oxidation barrier for ferromagnetic nano structures, the XAS of Py strips at the Fe and Ni edge were recorded after \SI{7}{days} in ambient conditions (7d), after \SI{84}{days} in ambient conditions (84d) and after an additional \SI{66}{min} on a hotplate at \SI{200}{\degreeCelsius} in ambient conditions (84d*).

\begin{figure}[htbp]
	\includegraphics[scale=1]{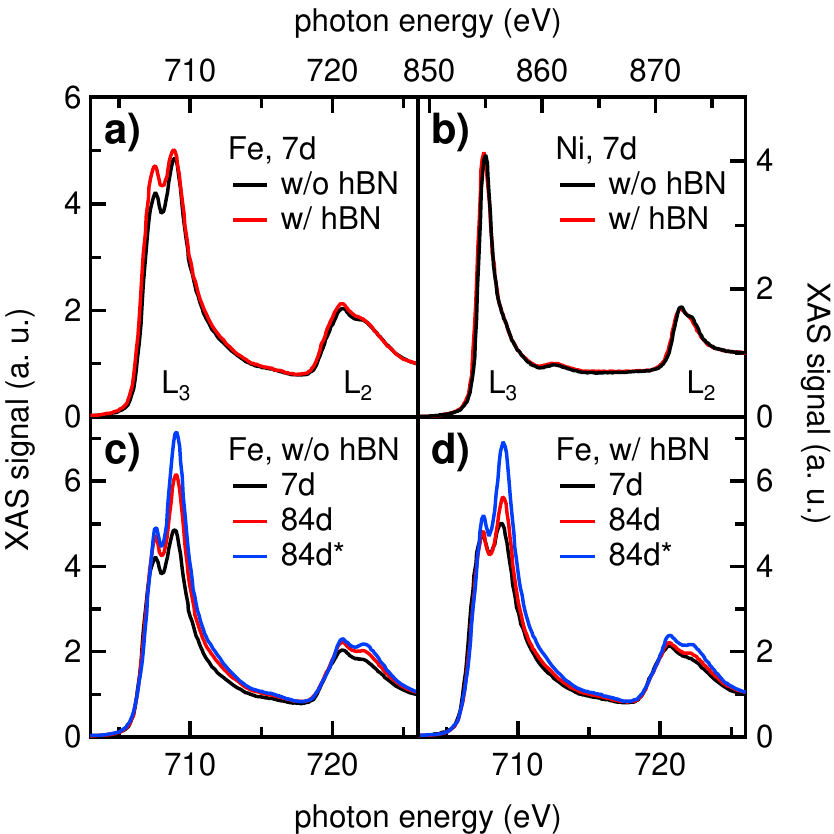}
	\caption{\label{fig:Py_spectra}\textbf{XAS spectra of Py strips}. \textbf{a)} and \textbf{b)} show XAS of Py strips at the Fe and the Ni edge, respectively. A direct comparison of Py strips covered with hBN (red) and uncovered regions (black) are shown after storing the samples for 7 days at ambient conditions. At the Fe-edge (\textbf{a)}), a pronounced difference in the spectra at the L$_3$ edge is observed. At the Ni-edge (\textbf{b)}) no pronounced difference is observed. Temporal evolution of the Fe-edge is shown in \textbf{c)} for an hBN covered region and in \textbf{d)} for an uncovered region.}
\end{figure}

A direct comparison of the XAS of a covered (w/ hBN) and an uncovered (w/o hBN) Py strip at the Fe edge is shown in Fig.~\ref{fig:Py_spectra}~\textbf{a)}. The smaller iron peak at the L$_3$ edge for the uncovered region indicates that there is a thicker oxide layer on top of the uncovered region (black curve) compared to the hBN covered (red). A similar, but smaller effect is also observed at the L$_2$ edge of the iron. At the Ni edge, there is no noticeable difference between the two different regions and no sign of oxidation, see Fig.~\ref{fig:Py_spectra}~\textbf{b)}. Altogether, it is clear that there is a significant difference in iron oxidation in Py strips between hBN covered and uncovered regions.

The evolution of the XAS at the Fe edge with time is shown in Fig.~\ref{fig:Py_spectra}~\textbf{c)} for an uncovered region and in \textbf{d)} for a hBN covered region. It is clear that in both cases the oxide peak grows with longer oxygen exposure time. Furthermore, the L$_2$ resonance of Fe starts to split into two peaks as well, indicating further oxidation. However, it is also clear that the oxide peak of the hBN covered region grows slower, especially for the XAS at 84d.

To quantify the amount of oxidation for the different samples we modelled the XAS signal and fitted the measured spectra. For that we assumed that the metallic ferromagnet is covered with a layer of oxide on top. In the case of Py, we treated the iron and the nickel individually. This is justified by the relatively small Fe content of the Py that favours individual oxidation of the elements as supported by our data. In short, oxidised Fe is found to co-exist with metallic Ni in agreement with the higher oxygen affinity of Fe compared to Ni \cite{1971_Reed}.

\begin{figure}[htbp]
	\includegraphics[scale=1]{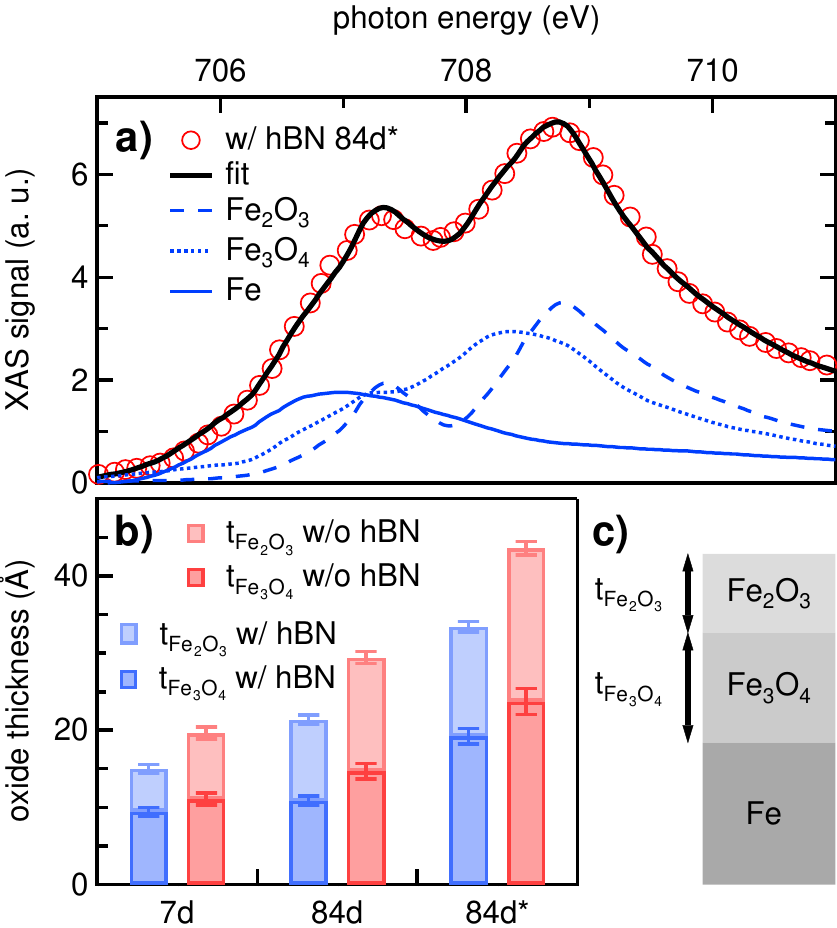}
	\caption{\label{fig:fit_Fe}\textbf{Fe oxide thickness of Py strips}. The XAS at the Fe-edge for an hBN covered region was fitted as a superposition of Fe, Fe$_3$O$_4$ and Fe$_2$O$_3$ spectra to extract the corresponding oxide thicknesses $t_{\mathrm{Fe}_3\mathrm{O}_4}$ and $t_{\mathrm{Fe}_2\mathrm{O}_3}$. The individual components are shown in blue. In \textbf{b)} the extracted oxide thicknesses are shown as a function of oxygen exposure time. In \textbf{c)} the model structure that was used to fit the XAS is depicted.}
\end{figure}

In the case of iron, a layer of Fe$_2$O$_3$ with a thickness $t_{\mathrm{Fe}_2\mathrm{O}_3}$ atop a layer of Fe$_3$O$_4$ with a thickness of $t_{\mathrm{Fe}_3\mathrm{O}_4}$ atop a Fe layer of infinite thickness was assumed. FeO was neglected since it is only stable under conditions of limited oxygen availability \citep{2001_Regan, 2014_Vaz}. Since Fe$_2$O$_3$ is the higher oxidised state of iron, we assume that the best model structure is given in Fig.~\ref{fig:fit_Fe}~\textbf{c)} where Fe$_2$O$_3$ is the topmost layer. Details about the fitting procedure can be found in the supporting information. We were unable to reasonably fit the measured data with a single oxide layer only. In Fig.~\ref{fig:fit_Fe}~\textbf{a)} we show a fit to a measured XAS, showing excellent agreement. By fitting the XAS for all different conditions we are able to extract the individual oxide thicknesses, which are shown in Fig.~\ref{fig:fit_Fe}~\textbf{b)}. The error bars correspond to the standard deviations obtained from least square fits. The statistical error (variance between different regions) is smaller than the error obtained from the fitting and therefore the statistical contribution was neglected.

Our results show that the oxide is always significantly thinner for the hBN covered regions compared to uncovered regions, see Fig.~\ref{fig:fit_Fe}~\textbf{b)}. It is also obvious that both oxide layers increase in thickness with longer oxygen exposure time. This is the case for hBN covered and uncovered regions, but more significant for the uncovered regions. Whereas $t_{\mathrm{Fe}_3\mathrm{O}_4}$ is similar for hBN covered and uncovered regions, the Fe$_2$O$_3$ layer is much thicker for the uncovered regions at \SI{7}{d}. As the Py strips are further exposed to air (84 d), mainly $t_{\mathrm{Fe}_2\mathrm{O}_3}$ increases with moderate changes in $t_{\mathrm{Fe}_3\mathrm{O}_4}$. The oxidation is promoted by putting the sample on a hotplate at \SI{200}{\degreeCelsius} as indicated by the increase in the oxide layer thickness and by a modification in the relative weight of the two oxides.

\begin{figure}[htbp]
	\includegraphics[scale=1]{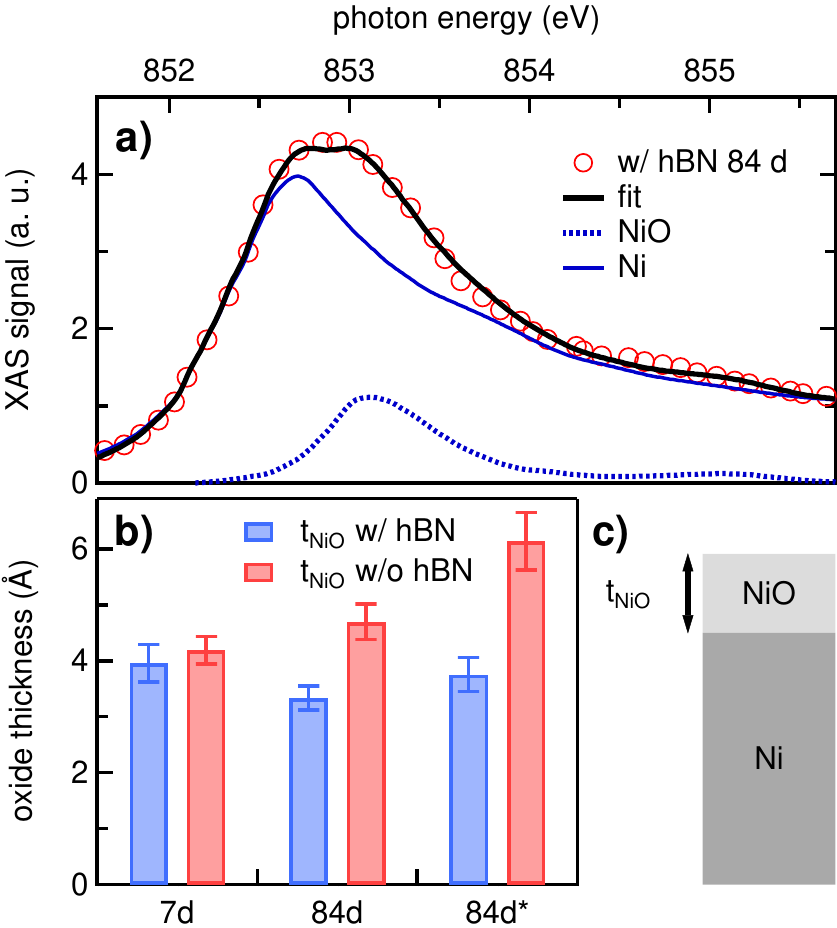}
	\caption{\label{fig:fit_Ni}\textbf{Ni oxide thickness of Py strips}. The XAS at the Ni-edge for an hBN covered region was fitted as a superposition of Ni and NiO spectra to extract the oxide thickness t$_{\mathrm{Ni}\mathrm{O}}$. The individual components are shown in blue. In \textbf{b)} the extracted oxide thicknesses are shown as a function of oxygen exposure time. \textbf{c)} Model structure used to fit the XAS spectra of the Ni-edge.}
\end{figure}

Similar to the Fe edge, the Ni L$_3$ edge was used to extract the thickness of the NiO. Higher oxidation states of nickel were neglected since they only form at higher temperatures \cite{2013_Liu}. In Fig.~\ref{fig:fit_Ni}~\textbf{a)} we show a fit to a measured XAS after \SI{84}{d} in ambient conditions, showing also excellent agreement. The individual components are also shown in blue. The metallic Ni is dominating, indicating a thin NiO layer. By fitting the XAS for all different conditions, we were able to extract the individual oxide thicknesses, which are shown in Fig.~\ref{fig:fit_Ni}~\textbf{b)}.

Initially, one finds no difference between hBN covered and uncovered regions in the oxidation of Ni, see Fig.~\ref{fig:fit_Ni}~\textbf{b)} at \SI{7}{d}. Upon further oxygen exposure, the thickness of the NiO stays the same within the error bars for hBN covered regions. In contrast to that, Ni oxidises further in the the case of uncovered Py strips and $t_{\mathrm{NiO}}$ increases by \SI{50}{\%} to \SI{6.1(5)}{\angstrom}.

\begin{figure}[htbp]
	\includegraphics[scale=1]{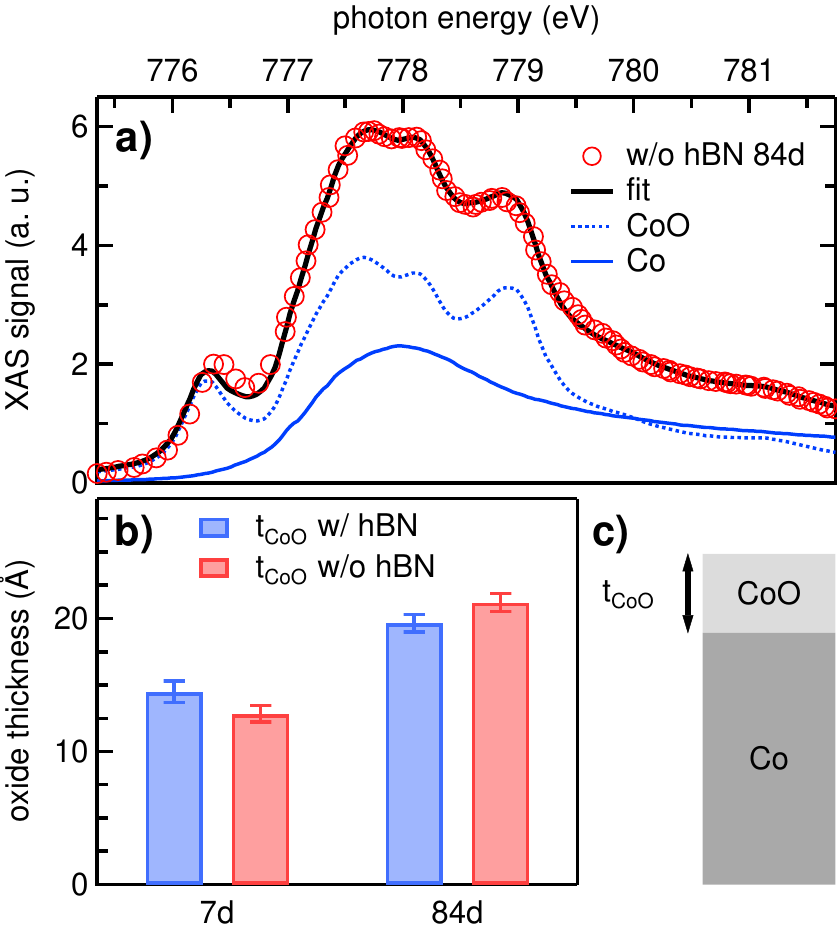}
	\caption{\label{fig:fit_Co}\textbf{Co oxide thickness in Co strips}. The XAS at the Co edge for an hBN covered region was fitted as a superposition of Co and CoO spectra to extract the oxide thickness t$_{\mathrm{Co}\mathrm{O}}$. The individual components are shown in blue. In \textbf{b)} the extracted oxide thicknesses are shown as a function of oxygen exposure time. Spectra were normalised with the I$_0$ counts from a mirror and not with a reference area. \textbf{c)} Model structure used to fit the XAS spectra of the Co-edge.}
\end{figure}

We have done a similar analysis for hBN covered and uncovered cobalt strips, where we extracted the thickness of the CoO ($t_{\mathrm{CoO}}$). Higher oxidation states of cobalt were neglected as in the Ni case. In Fig.~\ref{fig:fit_Co}~\textbf{a)} we show a fit to a measured XAS after \SI{84}{d} in ambient conditions, showing excellent agreement. By fitting the XAS for all different conditions we are able to extract the individual oxide thicknesses, which are shown in Fig.~\ref{fig:fit_Co}~\textbf{b)}.

Although it is not obvious on first sight that Co strips covered with hBN oxidise more slowly, the absolute increase of \SI{5.1(10)}{\angstrom} is significantly smaller than the increase of \SI{8.4(9)}{\angstrom} for the uncovered region. It is important to note that markers on the samples allowed us to look at the same location in different measurements (e.g. 7d and 84d). 

Table \ref{tab:Fe} and \ref{tab:NiCo} show an overview of the extracted oxide thicknesses for the different elements and samples. The difference ($\Delta$) is calculated by subtracting the oxide thickness with hBN from the oxide thickness without hBN.

\begin{figure}[htbp]
	\includegraphics[scale=1]{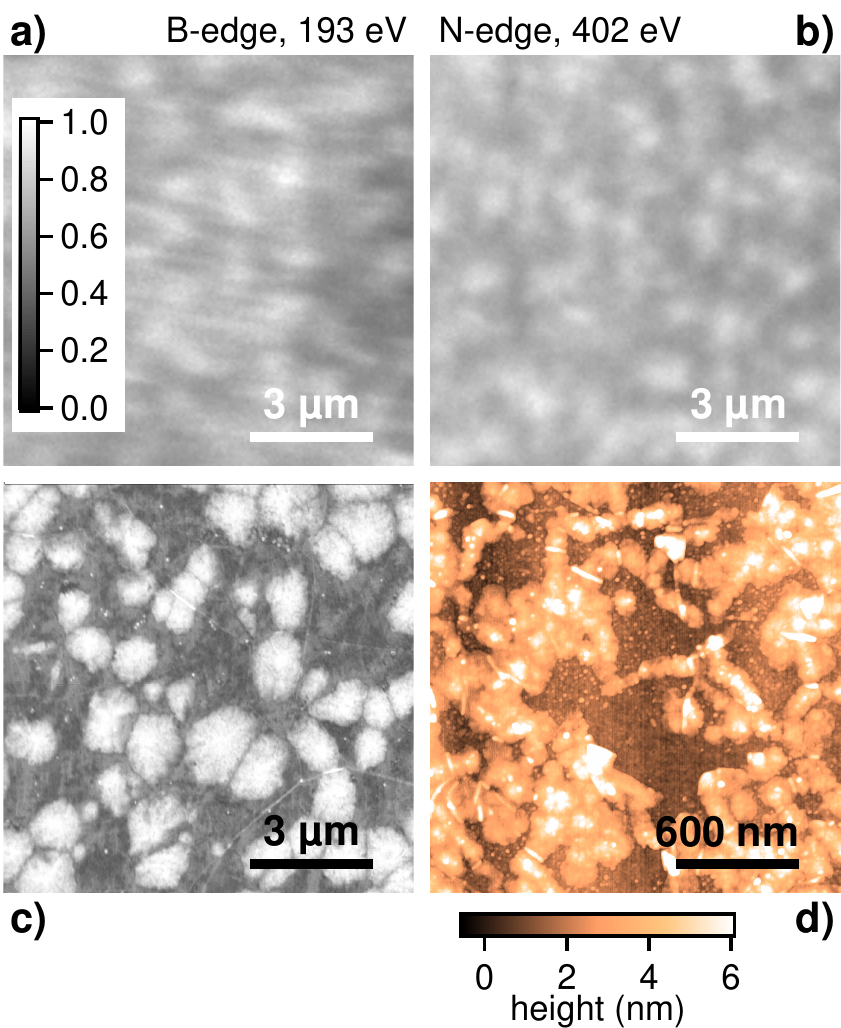}
	\caption{\label{fig:hBN}\textbf{hBN characterisation}. PEEM images of the Py strips covered with hBN at the K-edge of boron (\textbf{a)}) and at the K-edge of nitrogen (\textbf{b)}). Both images are edge / pre edge images and normalized and the scale bar in a) also applies for b). At both edges, the inhomogeneous nature of the hBN is clearly visible. The better image quality in \textbf{b)} compared to \textbf{a)} is due to a better adjusted electron optics in the PEEM. The spatial variation in the PEEM images correlates well with the structural features observed in the SEM image (\textbf{c)}), which was recorded with an acceleration voltage of \SI{2}{kV}. In \textbf{d)}, an AFM scan of an hBN layer transferred to a Si/SiO$_2$ wafer is shown over a range of \SI{2}{micrometer}, also showing a highly inhomogeneous hBN layer, even on a shorter length scale.}
\end{figure}

Since the hBN is a crucial part of this study, we characterised the hBN with several techniques, including PEEM, SEM and atomic force microscopy (AFM). PEEM images of hBN transferred on top of ferromagnetic strips are shown in Fig.~\ref{fig:hBN} recorded at the boron K-edge (\textbf{a)}) and at the nitrogen K-edge (\textbf{b)}). Both images are edge / pre edge images where the image at the edge is divided by an image at the pre edge. This corrects for unwanted contributions from the detectors stage and it also corrects for surface effects such that the observed contrast is a pure material contrast. Furthermore, the intensity has been normalized. In both images, spatial intensity variations are clearly visible indicating an inhomogeneous hBN layer, largely varying in thickness. Similar structures are also observed in the SEM image (see Fig. \ref{fig:hBN}~\textbf{c)}). This SEM image was taken on the as-received hBN on copper foil and therefore we conclude that this inhomogenities are not introduced by the transfer, but are rather a sign of a low quality hBN layer. AFM characterisation of an hBN layer transferred to a Si/SiO$_2$ wafer shown in \textbf{d)} reveal height variations of around \SI{2}{nm} which are far from the thickness of a single layer (\SI{\approx3}{\angstrom}). From the observations, we conclude that the hBN layers have not only multilayer patches but may also have holes and cracks.

\begin{table*}[hbtp]
	\centering
	\renewcommand{\arraystretch}{1.3}
	\caption{\label{tab:Fe} Overview of the extracted iron oxide thicknesses for regions without hBN (w/o hBN), with hBN (w/hBN) and the absolute difference between these regions ($\Delta$).}
    \begin{tabularx}{0.8\textwidth}{llXXXlXXX}
    \toprule[1.5pt]
        ~    & ~  & ~            & $t_{\mathrm{Fe}_2\mathrm{O}_3}$ (\si{\angstrom})    & ~      & ~          & ~              & $t_{\mathrm{Fe}_3\mathrm{O}_4}$ (\si{\angstrom})   & ~                \\ 
        ~  & ~  & w/o hBN           & w/ hBN            & $\Delta$ & ~        & w/o hBN          & w/ hBN            & $\Delta$
        \\ \hline
        7d & ~  & \num{8.5(8)} & \num{5.6(6)} & \num{-3.0(13)}& ~ & \num{11.1(8)}  & \num{9.5(6)} & \num{-1.6(14)} \\ 
        84d & ~ & \num{14.7(8)} & \num{10.5(6)} & \num{-4.2(14)}& ~ & \num{14.7(10)} & \num{10.8(6)} & \num{-3.9(16)} \\ 
        84d* & ~ & \num{19.8(9)} & \num{14.2(7)} & \num{-5.6(15)}& ~ & \num{23.8(17)} & \num{19.2(9)} & \num{-4.6(27)} \\
    \bottomrule[1.5pt]
    \end{tabularx}
\end{table*}

\begin{table*}
	\centering
	\renewcommand{\arraystretch}{1.3}
	\caption{\label{tab:NiCo} Overview of the extracted nickel and cobalt oxide thicknesses  for regions without hBN (w/o hBN), with hBN (w/hBN) and the absolute difference between these regions ($\Delta$).}
    \begin{tabularx}{0.8\textwidth}{llXXXlXXX}
    \toprule[1.5pt]
        ~    & ~ & ~             & $t_{\mathrm{NiO}}$ (\si{\angstrom})           & ~  & ~               & ~              & $t_{\mathrm{CoO}}$ (\si{\angstrom})            & ~                \\ 
        ~  & ~  & w/o hBN           & w/ hBN            & $\Delta$ & ~        & w/o hBN          & w/ hBN            & $\Delta$    
        \\ \hline
        7d & ~  & \num{4.2(3)} & \num{4.0(3)} & \num{-0.2(6)}& ~ & \num{12.8(6)} & \num{14.5(8)} & \num{1.7(15)}  \\ 
        84d & ~ & \num{4.7(3)} & \num{3.3(2)} & \num{-1.4(5)}& ~   & \num{21.2(7)} & \num{19.6(6)} & \num{-1.6(13)} \\ 
        84d* & ~ & \num{6.1(5)} & \num{3.8(3)} & \num{-2.4(8)}& ~   & ~              & ~              & ~                \\
    \bottomrule[1.5pt]
    \end{tabularx}
\end{table*}

\section{Discussion}
\label{sec:discussion}
Summarizing the data presented above, we can clearly say that there is a difference in oxidation of Fe and Ni for hBN covered and uncovered strips, namely that the hBN covered strips are less oxidised. This is most obvious when comparing the increase in oxide thickness from \SI{7}{d} to \SI{84}{d} oxygen exposure time.

Ni in close proximity to Fe clearly oxidises slowly and only a very thin oxide layer forms at the top. This can easily be explained by the fact that Fe oxidises first as it has a higher oxygen affinity than the Ni \cite{1971_Reed}. Ni will then only start to oxidise if all iron in close proximity is fully oxidised.

The thin oxide layer already present at \SI{7}{d} for all samples is partially due to the fabrication process used here. After evaporation in vacuum, the strips are brought to ambient conditions for around \SI{30}{min} for lift-off. This is already enough for the ferromagnetic materials to oxidise to a certain depth. 

Treating the Py as pure Fe for the fitting of the Fe L$_3$ edge probably underestimates the thicknesses of the oxides since only \SI{20}{\%} of the atoms in the Py are iron atoms. This might lead to extracted values that are smaller than in reality, but the change over time is still captured well. At the absorption edge of Ni, no Fe related features in absorption are expected, nor vice versa, since the energies of the photons are very different. Furthermore, the electron escape depth is similar for Fe ($\lambda_{Fe}$~=~\SI{15}{\angstrom}) and Ni ($\lambda_{Ni}$~=~\SI{22}{\angstrom}).

In the case of the Co strips, the protection of the hBN against oxidation is not obvious. The behaviour observed is attributed to the low quality of the hBN layers.

During the heat treatment at \SI{200}{\degreeCelsius} in ambient conditions, the relative weight of the two iron oxides is most probably changed due to different activation energy in their formation. The ratio of $t_{\mathrm{Fe}_2\mathrm{O}_3} / t_{\mathrm{Fe}_3\mathrm{O}_4}$ decreased with the heat treatment indicating a shift towards Fe$_3$O$_4$. In addition, an increased temperature also leads to a faster oxygen diffusion within the Py strip. This might be the reason why there is more Fe$_3$O$_4$ after the hotplate treatment.

We have also seen an increase in the oxide thickness for the Fe and Co in the hBN covered regions, although it is less pronounced. A defective layer of hBN with some holes or cracks would surely allow for some oxygen diffusion through the layer. In addition, grain boundaries could also allow oxygen diffusion through the layer. These two issues are related to the quality of the hBN and could be minimized by a higher quality hBN (e.g. more homogeneous layer and larger crystals). The mechanical transfer of the hBN on top of the ferromagnetic strips could also lead to cracks along the edges of the strips. This is possible since the height of the strips (\SI{30}{nm}) greatly exceeds the hBN thickness and therefore the hBN could rupture along this step edge. Heating the sample to elevated temperature could even promote this rupturing since the different materials involved have different thermal expansion coefficients. Moreover, oxygen diffusion along the \mbox{hBN / SiO$_2$} interface is possible, even though long distances have to be overcome. A defective hBN layer would facilitate the diffusion along the interface since the distances can be orders of magnitude smaller.

To sum up, the quality and the ex-situ transfer of hBN on top of the ferromagnetic material limits the effective protection of the hBN. This could possibly be solved by an in-situ transfer of high quality hBN onto pre-patterned and pre-cleaned ferromagnetic electrodes (e.g. transfer in an UHV system with sputter cleaned ferromagnets). Even more attractive is the direct CVD growth of hBN on predefined ferromagnetic strips. Recently, large single crystal CVD growth of hBN on a Fe catalyst has been achieved \cite{2015_Caneva}, paving the way for hBN protected iron strips.

\section{Conclusion}
\label{sec:Conclusion}
We conclude that the hBN slows down the oxidation of permalloy nanostructures. However, the ex-situ transfer of hBN has its limitations as the ferromagnetic materials are shortly exposed to air which is enough to oxides ferromagnetic surfaces several atomic layers deep. These mostly antiferromagnetic oxide layers might substantially decrease the spin-polarization of the electrons in electrical spin injection experiments. Most probably, in-situ grown hBN on ferromagnetic contacts could be a better solution for spin injectors into non-magnetic materials such as graphene as it most probably would protect the ferromagnetic material from oxidation. In addition this hBN coating could serve as a perfect tunnel barrier for electrical spin injection. Further investigations with higher quality hBN layers need to be done to clarify the performance of hBN as a potential oxidation barrier for ferromagnetic nanostructures.

\section{Acknowledgements}
\label{sec:Acknowledgements}
The authors gratefully thank Gulibusitan Abulizi for the atomic force microscopy picture of the hBN. We also thank Armin Kleibert and Frithjof Nolting for fruitful discussions. Part of this work was performed at the Surface/Interface: Microscopy (SIM) beam line of the Swiss Light Source, Paul Scherrer Institut, Villigen, Switzerland. This work was further funded by the Swiss National Science Foundation, the Swiss Nanoscience Institute, the Swiss NCCR QSIT, the ERC Advanced Investigator Grant QUEST and the EU Flagship project graphene.


\bibliographystyle{unsrt}
\bibliography{literature_manuscript}

\end{document}


\title{Supporting information: Role of hexagonal boron nitride in protecting ferromagnetic nanostructures from oxidation}


\author{Simon Zihlmann}
\affiliation{Department of Physics, University of Basel, Klingelbergstrasse 82, CH-4056 Basel, Switzerland}

\author{P\'eter Makk}
\email{peter.makk@unibas.ch}
\affiliation{Department of Physics, University of Basel, Klingelbergstrasse 82, CH-4056 Basel, Switzerland}

\author{C. A. F. Vaz}
\affiliation{Swiss Light Source, Paul Scherrer Institut, CH-5232 Villigen PSI, Switzerland}

\author{Christian Sch\"onenberger}
\affiliation{Department of Physics, University of Basel, Klingelbergstrasse 82, CH-4056 Basel, Switzerland}

\date{\today}
%

\maketitle

\section{Fabrication details}
\label{sec:FabricationDetails}
First, ferromagnetic strips were fabricated on highly p doped silicon substrate (only native oxide). Standard e-beam lithography with a positive resist (ZEP) was used to pattern the ferromagnetic strips. Py and Co were evaporated from an e-gun target in a UHV system at a pressure of \SI{\leq5e-10}{mbar}. A total film thickness of \SI{30}{nm} was deposited at a deposition rate of \SI{0.3}{\angstrom/s}. Lift-off was performed in N-Methyl-2-Pyrrolidon (NMP) at \SI{70}{\degreeCelsius} followed by an acetone and a 2-Propanol (IPA) rinse. Directly after lift-off, a bilayer (BL) hexagonal boron nitride (hBN) was transferred on top of the samples such that half of it was covered by hBN. Chemical vapour deposited (CVD) hBN grown on copper foil was purchased from Graphene Supermarket. A thick PMMA layer was spin-coated onto the as received copper foil with hBN. The copper substrate was then etched in a \SI{0.35}{mmol\per L} ammoniumpersulfate solution, leaving an hBN layer with the supporting PMMA layer floating on water. This layer was then transferred onto another piece of hBN covered copper substrate to obtain a BL hBN. Now, a PDMS stamp was added to the PMMA support layer to increase the stability and to make it easier to handle. Again, the Cu substrate was etched in an ammoniumpersulfate solution. After thoroughly washing with water and IPA and drying, the BL hBN was transferred in a dry process onto the ferromagnetic films/strips. Before removing the supporting PMMA and PDMS with a hot acetone bath for \SI{30}{min}, the sample was cured at \SI{180}{\degreeCelsius} for \SI{3}{min} to relax the PMMA and to enhance the adhesion of the transferred hBN.

\section{Fitting procedure}
\label{sec:FittingProcedure}

The oxide thickness of the ferromagnetic strips was extracted using a procedure introduced by Regan et al. \cite{2001_Regan}, which models the total electron yield (TEY) of an oxide / metal sandwich. In the following, the derivation is given for a stack consisting of Fe with an Fe$_2$O$_3$ layer atop a Fe$_3$O$_4$ layer atop the metallic iron, but it is analogous for Ni and Co with their oxides. The total electron yield $dN_\mathrm{Fe_2O_3}$ from a layer of Fe$_2$O$_3$ of thickness $dz$ at depth $z$ is given by
\begin{equation}
	\label{eq:dTEY_Fe2O3}
	\begin{split}
	dN_\mathrm{Fe_2O_3} = I_0 \cdot e^{-z\mu_{Fe_2O_3}(E)} \cdot \mu_{Fe_2O_3}(E) \cdot G_{Fe_2O_3}(E) \cdot e^{-z/\lambda_{Fe_2O_3}} \cdot dz
	\end{split}
\end{equation}
assuming normal incidence of the photons.
Here $I_0$ is the photon flux, $\mu_{Fe_2O_3} (E)$ is the absorption coefficient representing the probability of a photon being absorbed, $G_{Fe_2O_3}(E)$ is the number of electrons created per absorbed photon and $\lambda_{Fe_2O_3}$ is the electron escape depth. The first exponential factor gives the probability of a photon reaching depth $z$ and the second exponential factor gives the probability of a photo electron created at depth $z$ to escape to the surface. Integration of equation~\ref{eq:dTEY_Fe2O3} from 0 (surface) to depth $t_{Fe_2O_3}$ gives the total electron yield of a Fe$_2$O$_3$ layer with thickness $t_{Fe_2O_3}$ as follows:
\begin{equation}
	\label{eq:TEY_Fe2O3}
	\begin{split}
		N_\mathrm{Fe_2O_3} = \frac{I_0 \cdot G_{Fe_2O_3}}{1 + \frac{1}{\lambda_{Fe_2O_3} \cdot \mu_{Fe_2O_3}}} \cdot e^{-t_{Fe_2O_3}\left(\mu_{Fe_2O_3} + 1/\lambda_{Fe_2O_3} \right)}.
	\end{split}
\end{equation}
Here the energy dependence is not explicitly indicated.

Similarly, one can derive the total electron yield of the Fe$_3$O$_4$ layer below, ranging from $z_1 = t_{Fe_2O_3}$ to $z_2 = t_{Fe_3O_4}$, as follows:
\begin{equation}
	\label{eq:TEY_Fe3O4}
	\begin{split}
		N_\mathrm{Fe_3O_4} = \frac{I_0 \cdot G_{Fe_3O_4}}{1 + \frac{1}{\lambda_{Fe_3O_4} \cdot \mu_{Fe_3O_4}}} \cdot e^{-t_{Fe_2O_3}\left(\mu_{Fe_2O_3} + 1/\lambda_{Fe_2O_3}\right)} \cdot \left[ 1 - e^{-\left(t_{Fe_3O_4} - t_{Fe_2O_3}\right)\cdot\left(\mu_{Fe_3O_4} + 1/\lambda_{Fe_3O_4} \right)} \right].
	\end{split}
\end{equation}
Here the fist exponential accounts for the absorption and electron escape losses of the Fe$_2$O$_3$ layer.

The total electron yield of the metallic iron (infinite thickness) below these two oxide layers is then given by:
\begin{equation}
	\label{eq:TEY_Fe}
	\begin{split}
		N_\mathrm{Fe} = \frac{I_0 \cdot G_{Fe}}{1 + \frac{1}{\lambda_{Fe} \cdot \mu_{Fe}}} \cdot e^{-t_{Fe_2O_3}\left(\mu_{Fe_2O_3} + 1/\lambda_{Fe_2O_3}\right)} \cdot  e^{-\left(t_{Fe_3O_4} - t_{Fe_2O_3}\right)\cdot\left(\mu_{Fe_3O_4} + 1/\lambda_{Fe_3O_4} \right)}.
	\end{split}
\end{equation}
The two exponential account for the absorption and electron escape losses due to the two oxide layers.

The total electron yield of a Fe$_2$O$_3$ / Fe$_3$O$_4$ / Fe sandwich as shown in Fig.~3~\textbf{c)} in the main text is then given by the sum of equation~\ref{eq:TEY_Fe2O3}, \ref{eq:TEY_Fe3O4} and \ref{eq:TEY_Fe}. For the fitting, it was assumed that $G_{Fe_2O_3}(E) = G_{Fe_3O_4}(E) = G_{Fe}(E) = const$ (neglecting the difference of the electron yield for different materials and their energy dependence). This results in an overall scaling factor accounting for the photon flux and electron yield. Hence, only $t_{Fe_2O_3}$ and $t_{Fe_3O_4}$ determine the relative weight of the different chemical species involved and the fit is simply speaking a linear combination of the reference spectra of $\mu_{Fe_2O_3}$, $\mu_{Fe_3O_4}$ and $\mu_{Fe}$, which were all taken from Ref. \cite{2001_Regan}. For the electron escape depths we used values from the literature: $\lambda_{Fe_2O_3}$~=~\SI{35}{\angstrom}~\citep{1999_Nakajima}, $\lambda_{Fe_3O_4}$~=~\SI{50}{\angstrom}~\citep{2000_Gota}, $\lambda_{Fe}$~=~\SI{15}{\angstrom}~\citep{1998_Chakarian}, $\lambda_{Co}$~=~\SI{22}{\angstrom}~\cite{1999_Nakajima, 1998_Chakarian}, $\lambda_{Ni}$~=~\SI{22}{\angstrom}~\cite{1999_Nakajima, 1998_Chakarian}. To our knowledge, the electron escape depths for CoO and for NiO has not been determined and therefore we used a value of \SI{30}{\angstrom}, as suggested by Regan et al. \citep{2001_Regan}.


\bibliographystyle{unsrt}
\bibliography{literature_SOM}